\documentclass[letter]{aa}
\usepackage{orcidlink}
\usepackage{graphicx}
\usepackage{txfonts}
\usepackage{amsmath,amssymb}
\usepackage{microtype}
\usepackage{hyperref}
\usepackage[nameinlink,capitalize]{cleveref}
\usepackage{csquotes}
\usepackage[switch]{lineno}

\newcommand{\Msun}{M_{\odot}}

\newcommand{\Mgas}{M_{\rm gas}}
\newcommand{\Msix}{M_{\rm gas,6}}
\newcommand{\Rc}{R_{\rm c}}
\newcommand{\Rblow}{R_{\rm blow}}
\newcommand{\fmech}{f_{\rm mech}}

\newcommand{\Mstar}{M_{*}}

\newcommand{\xid}{\xi_{d}}
\newcommand{\xidzero}{\xi_{d,0}}

\begin{document}

\titlerunning{Blue Monsters: missing dust and mechanical blowout}
\authorrunning{Martínez-González et al.}

\title{On the missing dust of super-early galaxies}
\subtitle{Supernova blowout and gas–dust venting in Blue Monsters}

\author{
  Sergio Mart\'{\i}nez-Gonz\'alez
  \thanks{Corresponding author. E-mail: sergiomtz@inaoep.mx}%
  \inst{1}\orcidlink{0000-0002-4371-3823}
  \and
  Santiago Jim\'enez
  \inst{2}\orcidlink{0000-0003-2808-3146}
  \and
  Casiana Mu\~{n}oz-Tu\~{n}\'on
  \inst{3,4}\orcidlink{0000-0001-8876-4563}
}

\institute{
  Instituto Nacional de Astrof\'\i sica, \'Optica y Electr\'onica, AP 51, 72000 Puebla, M\'exico
\and
  Astronomical Institute of the Czech Academy of Sciences,  Bo\v{c}n\'\i\ II 1401/1, 141 00 Praha 4, Czech Republic
\and
  Instituto de Astrof\'isica de Canarias, E 38200 La Laguna, Tenerife, Spain
\and 
  Departamento de Astrof\'isica, Universidad de La Laguna, E 38205 La Laguna, Tenerife, Spain
}

\abstract
{A subset of very young, super-early galaxies at $z\gtrsim10$, often termed Blue Monsters, shows extremely blue UV continua and faint far-IR emission, which could imply much less dust than expected from standard enrichment scenarios.}
{Understanding the possible reason behind the apparent absence of dust in the Blue Monsters. For that, we show how clustered supernovae drive mechanical blowout in stratified, self-gravitating clouds by combining full 3-D hydrodynamical dust-survival yields with 3-D thin-shell scalings, and predict the retained dust-to-stellar mass ratio at the cluster scale and the corresponding galaxy-integrated value.}
{We take the net dust yield per unit stellar mass from existing 3-D hydrodynamical studies of young stellar clusters with sequential supernovae, and we set the blowout radius as a function of gas concentration using established 3-D thin-shell scalings. Assuming gas–dust coupling across the blowout boundary, the retained dust-to-stellar ratio accounts for the fraction of supernovae that remain confined versus those that vent mechanically.}
{Across typical cluster masses, sizes, and cloud-scale star formation efficiencies, mechanical venting removes a large share of gas and dust. The retained dust-to-stellar mass ratio is lowered by about half to two orders of magnitude relative to the supernova net dust yield. The outcome depends mainly on gas concentration and only weakly on metallicity, so it remains effective at low $Z$. After weighting by a Schechter cluster mass function and a Weibull core–radius distribution, the galaxy-integrated value falls in the same range inferred for spectroscopically-confirmed Blue Monsters.}
{Mechanical venting at the cluster scale can account for the very low dust fractions inferred for Blue Monsters without requiring extreme in situ destruction and without fine-tuning.}


\maketitle


\section{Introduction}
\label{intro}

Very young, compact super-early galaxies with blue UV slopes and weak far-IR emission require efficient removal or destruction of freshly condensed dust from star-forming regions, or a dust population with intrinsically low UV opacity (e.g., large supernova-produced grains) that can yield very blue continua \citep{Narayananetal2025}. These pathways are relevant for the population often termed \enquote{Blue Monsters}, whose extreme UV colors $\beta\simeq-2.5$ to $-2.6$ and very low extinction at $z\gtrsim10$ challenge standard dust and metal enrichment scenarios \citep{ArrabalHaroetal2023,Ziparoetal2023,Yoonetal2023,Bakxetal2023,Ferraraetal2025,Cullenetal2024,Zavalaetal2025}. In models without outflows, even allowing wide variations in dust yields and destruction, the predicted dust-to-stellar mass ratio $\xid \equiv M_d/\Mstar$ falls around $\log\xid \simeq -2.2$, whereas interpreting current Blue Monsters V-band attenuation implies $\log\xid \lesssim -4$ \citep{Ferraraetal2025}. In turn, this mismatch may point to a dynamical origin of transparency: feedback must clear grains faster than the young stellar population builds up, most efficiently along low-resistance paths in a clumpy medium. There, clustered winds and supernovae carve low-density corridors that drive mechanical blowout, venting hot gas and newly formed dust from their birth clouds.

Supernovae typically explode in clustered star-forming complexes embedded in non-uniform media \citep{Parizotetal2004}, where density asymmetries shape their evolution. Since mechanical blowout is most effective on cluster scales, its importance increases as star formation becomes more clustered at early times, and the resulting venting delays enrichment in star-forming regions and the chemical evolution of molecular clouds and galaxies \citep{Jimenezetal2019,Jimenezetal2021}. Empirically, the fraction of stars forming in compact stellar clusters rises with redshift and is consistent with dominance by $z\sim6$ \citep{Vanzellaetal2019}; this is supported by lensed pc-scale systems at $z=6.143$ \citep{Vanzellaetal2019}, $z_{\rm spec}=8.296$ \citep{Mowlaetal2024}, $z\approx10.2$ \citep{Adamoetal2024}, and $z=10.603$ \citep{Senchynaetal2024}, by the increase of cluster-formation efficiency with gas pressure \citep{Kruijssen2012,Adamoetal2015}, and by the rise of the star-formation-rate surface density from $z\sim4$ to $z\sim10$ \citep{Calabroetal2024}. At the galaxy scale, Blue Monsters are compact in the rest-UV, with half-light radii of order 0.1 to 1 kpc at $9\lesssim z\lesssim13$, as summarized by \citet[and references therein]{Somervilleetal2025}. This compactness shortens the path to the vertical scale height, allowing low-column channels from pc-scale blowout to remain open and vent gas and dust out of the galaxy \citep{TenorioTagle1996}.

Here we link full 3-D hydrodynamical simulations of supernova-driven dust injection and destruction in massive star clusters to thin-shell blowout scalings and translate them into predictions for the transparency of Blue Monsters. 
The structure of the paper is as follows: \Cref{sec:model} develops the analytic framework, coupling dust injection and survival in star clusters (subsection \ref{subsec:injection}) to mechanical venting from thin-shell blowout (subsection \ref{subsec:venting}); \Cref{sec:results} explores the parameter space, maps the retained dust-to-stellar mass ratio, and delineates the combinations of cluster masses, sizes, and cloud-scale star formation efficiency for which the galaxy-integrated $\log\langle\xid\rangle_{gal}$ settles around $-4$; \Cref{sec:limitations} presents the model limitations; and \Cref{sec:conclusions} summarizes the main results and their implications for Blue Monsters.

\section{Model}
\label{sec:model}

In order to assess the transparency of Blue Monsters, our model couples full 3-D hydrodynamical simulations of supernova-driven dust injection and destruction in massive star clusters with thin-shell blowout scalings.

\subsection{Dust injection and survival}
\label{subsec:injection}

The 3-D hydrodynamical simulations of \citet{MartinezGonzaletal2022} follow dust injection and survival from 186 pair-instability and 50 core-collapse supernovae in a young massive stellar cluster within a low-metallicity clumpy molecular cloud. They include, apart from the stellar and gaseous gravitational pull, turbulent motions and radiative cooling, explicit treatment of multiple dust processing channels: the passage of the supernova reverse shock through the ejecta, the injection of shocked stellar winds, and secondary forward shocks produced in sequential supernova explosions via thermal sputtering. The evolution is intrinsically non-spherical: off-center supernovae expand down local density gradients and into a wind-blown, corrugated shell, accelerating Rayleigh–Taylor growth, fragmenting the swept-up wall, and punching low-column density chimneys. The resulting shell is porous and highly anisotropic, providing preferential escape routes for gas and dust
\citep[for other processes see ][]{MartinezGonzalez2025a,MartinezGonzalez2025b}. These simulations adopt a Kroupa initial mass function \citep[IMF,][]{Kroupa2001} over $[0.01,120] \Msun$, metallicity $Z=2\times10^{-2}Z_\odot$, a star cluster mass of $5.6\times10^5 \Msun$, and the BoOST stellar evolutionary tracks \citep{Szecsietal2022}. For pair-instability supernovae a dust condensation efficiency of $3.2\%$ of the progenitor final mass was adopted \citep[][]{Cherchneff2010}; for core-collapse supernovae the injected dust mass was drawn from a normal distribution with mean $0.6 \Msun$ and standard deviation $0.1 \Msun$\footnote{In \citet{MartinezGonzaletal2022}, the choice of $0.6\pm0.1\,\Msun$ was motivated by supernova ejecta dust inventories: SN~1987A (\citealt{Matsuuraetal2015} report $\sim0.8\Msun$); Cas~A (\citealt{DeLoozeetal2017} find $\sim0.4$--$0.6\Msun$; \citealt{Bevanetal2017} infer up to $\sim1.1\Msun$); and G54.1+0.3 (\citealt{Rhoetal2018} report up to $\sim0.9\Msun$; \citealt{Temimetal2017} report a dust mass of $1.1\pm0.8\Msun$).}
. The fraction of the dust budget that survives the sequence of supernova explosions, is sustained at about $15$ to $20\%$, consistent with \citet{MartinezGonzalezetal2018} for type IIb and II-P supernovae from progenitors formed at solar metallicity in a young stellar cluster with mass $10^6\,\Msun$. We therefore adopt the scaling of \citet{MartinezGonzaletal2022}
\begin{equation}
  \xidzero \equiv \frac{M_d^{net}}{\Mstar}
  = \frac{1200\Msun}{5.6\times10^{5}\Msun}
  \simeq 2.14\times10^{-3}.
\end{equation}
Here $M_d^{net}$ denotes the dust mass that remains after shock processing; i.e., $\xidzero$ is the net supernova dust yield per unit stellar–cluster mass, corresponding to $\log\xidzero\simeq -2.669$.

\subsection{Mechanical venting}
\label{subsec:venting}

The venting of gas is taken into account following the thin-shell relations derived by \citet{Jimenezetal2019,Jimenezetal2021} which also include the gravitational pull of gas and star, turbulent pressure, and radiative cooling. They assume that gas and massive stars within individual stellar clusters follow Gaussian profiles of core radius $\Rc$. The decisive variable is the concentration

\begin{equation}
\mathcal{C} \equiv \frac{\Msix}{\Rc,_{\rm pc}}, \qquad \Msix \equiv \frac{\Mgas}{10^{6}\,\Msun},
\end{equation}

with $\Rc,_{\rm pc}$ in parsecs. The relations identify $\Rblow$, the radius at which the shell ceases to be pressure-confined, as

\begin{equation}
\frac{\Rblow}{\Rc}=
\begin{cases}
1.4246\left(\dfrac{\Msix}{\Rc,_{\rm pc}}\right)^{0.35}, & \dfrac{\Msix}{\Rc,_{\rm pc}}\le 2,\\[5pt]
1.5639\left(\dfrac{\Msix}{\Rc,_{\rm pc}}\right)^{0.20}, & \text{otherwise.}
\end{cases}
\label{eq:Rblow}
\end{equation}

Note $R_{\rm blow}$ depends on $M_{\rm gas}$ and $R_c$, not on metallicity $Z$, because in dense gas early supernova remnant evolution is dominated by free--free cooling \citep[see Fig.~7 in][]{Jimenezetal2021}.

Supernovae are distributed as a 3-D Gaussian of core radius $\Rc$. The enclosed supernova mass fraction is

\begin{equation}
F(<R)=\mathrm{erf}\!\left(\frac{R}{\sqrt{2}\Rc}\right)
-\sqrt{\frac{2}{\pi}}\frac{R}{\Rc}\exp\left(-\frac{R^{2}}{2\Rc^{2}}\right),
\end{equation}

which defines the mechanical venting fraction

\begin{equation}
\fmech(\Mgas,\Rc)=1-F(<\Rblow).
\end{equation}

Assuming gas–dust coupling, the retained dust-to-stellar mass ratio then follows

\begin{equation}
  \xid(\Mgas,\Rc)=\xidzero\,\bigl[1-\fmech(\Mgas,\Rc)\bigr].
\end{equation}

To map gas and stellar masses we use the scaling $\Mstar=\varepsilon_\star \Mgas$, where $\varepsilon_\star$ is the cloud-scale star formation efficiency. We obtain the galaxy-integrated dust-to-stellar mass ratio, $\langle\xi_d\rangle_{\rm gal}$, by averaging over the distributions of star cluster radii and masses. We use a Weibull distribution function for star cluster radii, mapping to the 3D–Gaussian core via $\Rc=R_{\rm eff}/\sqrt{2\ln 2}$. The radius distribution function, normalized to unity on $\Rc\in[0.5,20]$ pc, is $p_R(\Rc)=(k/\lambda_c)\,t^{k-1}e^{-t^k}/\mathcal{N}_R$, where $t=(\Rc-R_{0,c})/\lambda_c$, $\lambda_c=\lambda/\sqrt{2\ln 2}$, $R_{0,c}=R_0/\sqrt{2\ln 2}$, and $\mathcal{N}_R=e^{-t_{\min}^{k}}-e^{-t_{\max}^{k}}$ with $t_{\min}=(0.5-R_{0,c})/\lambda_c$ and $t_{\max}=(20-R_{0,c})/\lambda_c$; we use $k=2.22$, $\lambda=3.64$ pc, and $R_0=0.185$ pc \citep{BrownandGnedin2021}. We adopt a Schechter cluster mass function (CMF) with $\phi(M_\star)=A\,M_\star^{-2}e^{-M_\star/M_c}$ on $M_\star\in[10^3,10^7]\,M_\odot$ and $M_c=10^6\,M_\odot$ \citep[e.g.][]{Larsen2009}. The CMF weight is $W_M(M_\star)=e^{-M_\star/M_c}\,d\ln M_\star/\mathcal{N}_M$, where $\mathcal{N}_M=\int_{10^3}^{10^7} e^{-M/M_c}\,d\ln M$. Because $W_M$ and $p_R$ are normalized as defined above, $\mathcal{N}_M$, $\mathcal{N}_R$, and the CMF amplitude $A$ cancel. The galaxy–integrated ratio is thus

\begin{align}
\langle\xi_d\rangle_{\rm gal}
&= \int W_M(M_\star)\,d\ln M_\star \int p_R(\Rc)\,\xi_d\left(\frac{M_\star}{\varepsilon_\star},\Rc\right)\,d\Rc \nonumber\\
&= \frac{
\int_{\ln M_{\min}}^{\ln M_{\max}}
\int_{\Rc^{\min}}^{\Rc^{\max}}
M_\star^{2}\phi(M_\star)\,p_R(\Rc)\,\xi_d\left(\frac{M_\star}{\varepsilon_\star},\Rc\right)\,d\Rc\,d\ln M_\star}{
\int_{\ln M_{\min}}^{\ln M_{\max}}
\int_{\Rc^{\min}}^{\Rc^{\max}}
M_\star^{2}\phi(M_\star)\,p_R(\Rc)\,d\Rc\,d\ln M_\star } .
\end{align}

\section{Results}
\label{sec:results}

\begin{figure}[t]
\centering
\includegraphics[width=0.8\linewidth]{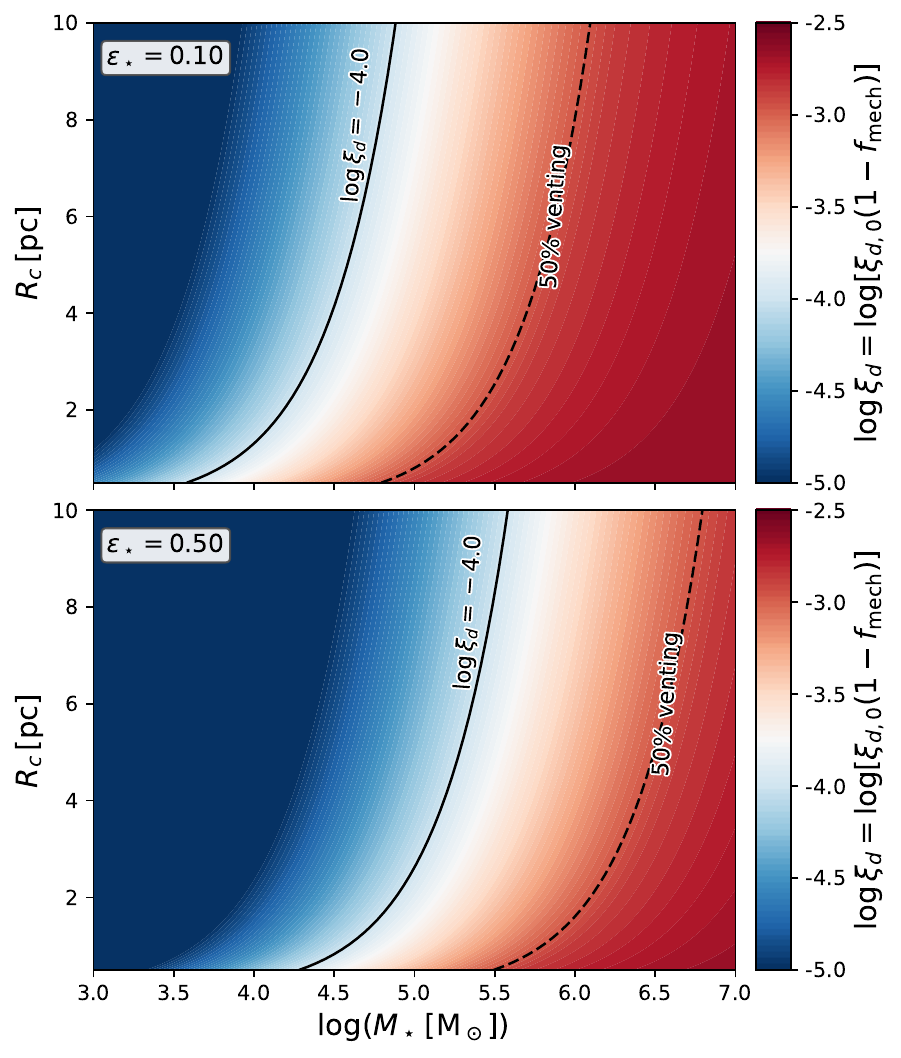}
\caption{Predicted retained dust-to-stellar mass ratio for star clusters as a function of stellar mass and star cluster core radius, shown for cloud-scale star formation efficiencies $\varepsilon_\star=0.10$ (upper panel), $0.50$ (lower panel). Colors encode $\log\xid=\log\!\big[\xidzero\,(1-\fmech)\big]$ with $\xidzero=2.14\times10^{-3}$. The dashed black curves mark $\fmech=0.5$, and the solid black contours display $\log\xid=-4$, the characteristic level inferred for Blue Monsters.}
\label{fig:1}
\end{figure}

Figure \ref{fig:1} maps the retained dust-to-stellar mass ratio $\log\xid$ for two cases of the cloud-scale star formation efficiency \citep[$\varepsilon_\star=0.10, 0.50$][]{Grudicetal2018}. The pattern follows the gas concentration $\mathcal{C}=\Msix/\Rc,_{\rm pc}$: at lower $\mathcal{C}$ the blowout radius grows relative to $\Rc$, a larger share of supernovae explode beyond $\Rblow$, and $\xid$ decreases. The dashed curves show $\fmech=0.5$ and the solid contours mark $\log\xid=-4$. Very massive and extremely compact star clusters with deep gravitational potentials maintain high $\mathcal{C}$ and suppress blowout, retaining more dust.

Since $\xid=\xidzero(1-\fmech)$ with $\log\xidzero\simeq -2.669$, reaching $\log\xid=-4$ needs $\fmech\gtrsim0.95$. The maps show that this level is attained across a broad domain of $(\Mstar,\Rc,\varepsilon_\star)$. As an example, for $\Rc\simeq5$ pc and $\varepsilon_\star=0.50$ the $\log\xid=-4$ contour appears near $\Mstar\approx10^5\,\Msun$. Extended clusters reach low $\xid$ at higher $\Mstar$ than compact ones because $\mathcal{C}$ is smaller.

\begin{figure}
  \centering
  \includegraphics[width=\columnwidth]{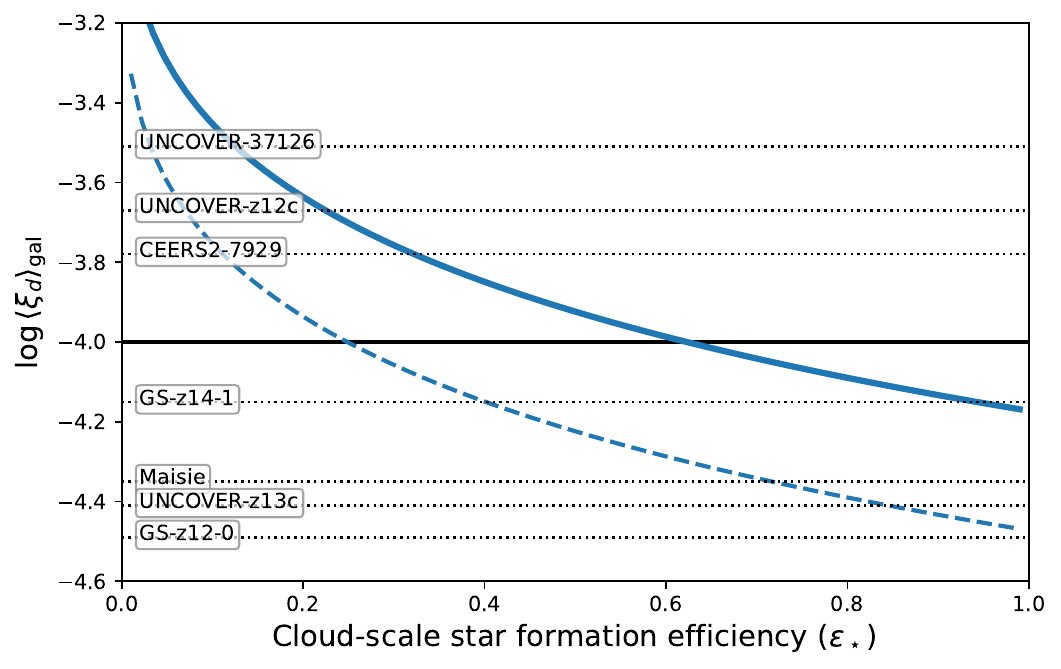}
  \caption{Galaxy-integrated dust-to-stellar mass ratio versus the cloud-scale star-formation efficiency. The solid blue curve corresponds to $\log\xidzero \simeq -2.669$, while the dashed blue curve corresponds to a factor-of-two reduction in $\xidzero$, consistent with either reduced supernova dust mass production or reduced dust survival. The horizontal solid line marks $\log\langle\xi_d\rangle_{\rm gal}=-4$. Dotted horizontal lines annotate $\log\langle\xi_d\rangle_{\rm gal}$ for a subset of spectroscopically-confirmed Blue Monsters as inferred by \citet{Ferraraetal2025}.}
  \label{fig:2}
\end{figure}

Figure \ref{fig:2} shows that, for $\varepsilon_\star=(0$–$1)$, the weighted $\log\langle\xid\rangle_{gal}$ lies around $-4$, without fine-tuning, and decreases monotonically with $\varepsilon_\star$. Dotted horizontal lines indicate the inferred, model-dependent $\log\langle\xid\rangle_{gal}$ for 
selected Blue Monsters, and they fall in the same range inferred from V-band attenuation in
\citet{Ferraraetal2025}. We note that the inferred \(\log\langle\xi_d\rangle_{\rm gal}\approx-4\) values depend on the assumed galaxy geometry and size, and on the dust model, and that line-of-sight extinction is sensitive to inclination and to the relative distribution of dust and stars, so low extinction does not by itself imply low total dust mass \citep[see][]{Chevallardetal2013,Sommovigoetal2025b}.

\section{Model limitations}
\label{sec:limitations}

The thin-shell scalings span a wider parameter space in gas concentration $\mathcal{C}=\Msix/\Rc,_{\rm pc}$ and in the core radius $\Rc$ (the same $\Rc$ is adopted for the stellar and gaseous components, both taken as Gaussian density profiles), while the 3-D hydrodynamical runs, although they follow hundreds of sequential supernovae with resolved shocked stellar winds and forward and reverse shocks, cover a narrower range of stellar cluster masses and $\Rc$. Using $\xidzero$ outside that mass range is therefore an extrapolation. 

Our current model omits both radiation pressure and radiative transfer. Radiation-pressure forces can decouple dust and gas, breaking the link between line-of-sight extinction and total dust mass \citep{Ziparoetal2023,Sommovigoetal2020,Menonetal2025}, while dust–star geometry and radiative transfer further modify the attenuation \citep{SalimandNarayanan2020,Narayananetal2018,Sommovigoetal2025b,DiMasciaetal2025,Matsumotoetal2025,Duboisetal2024,Draine2011}. In young star clusters, radiative and mechanical feedback act on comparable early timescales \citep{Rahneretal2017}. Radiation pressure can pre-clear low-column channels and accelerate early shell growth, which would increase the mechanical venting fraction $f_{\rm mech}$ at fixed $(M_{\rm gas}, R_c)$. Time-dependent models indicate that its dynamical dominance is brief and mainly precedes the supernova phase, after which winds and supernovae set the shell evolution \citep{SilichandTenorioTagle2013,MartinezGonzalezetal2014}. Because the supernova-condensed grains we track are produced after this phase, adding radiation pressure would likely strengthen the trend toward low $\langle\xid\rangle_{gal}$. Our mechanical-only $\xi_d$ should therefore be regarded as conservative upper limits.

We do not model IMF variations. A top-heavy IMF raises the specific supernova rate and UV momentum per unit stellar mass, boosting supernova dust production and early radiative and wind driving; this opens and sustains low-column channels, increases the mechanical venting fraction $f_{\rm mech}$ at fixed $(M_{\rm gas},R_c)$, and tends to lower the cloud-scale star-formation efficiency $\varepsilon_\star$ \citep{Menonetal2024}.

In contrast to the 3-D thin-shell simulations, the full 3-D hydrodynamical simulations place a Schuster stellar density distribution within a clumpy molecular cloud, so porosity and inhomogeneities are resolved.

Supernova–condensed dust yields, particularly at high redshift, remain highly uncertain; the adopted $\xidzero$ should be taken as an indicative normalization. On the positive side, the 15–20\% survival fractions used here come from on-the-fly grain processing within high-resolution 3-D hydrodynamical simulations of sequential supernovae, and are therefore comparatively robust \citep{MartinezGonzalezetal2018,MartinezGonzaletal2022}. The surviving dust is produced predominantly by off-centered supernovae \citep{MartinezGonzalezetal2018}. It follows that the fraction of surviving dust formed at $r>R_{\rm blow}$ is expected to be at least as large as the geometric fraction of supernovae exploding beyond $R_{\rm blow}$. Using $\xid=\xidzero\,[1-f_{\rm mech}]$ therefore yields an upper bound on the retained dust. We note most studies of reverse-shock-driven dust destruction, as reviewed by \citet{SchneiderandMaiolino2024}, assume uniform or phase-averaged media and derive efficiencies from Sedov-Taylor similarity solutions. In clustered supernovae, shocks transmitted into the wind-blown shell are strongly radiative, the shell is not overrun, and no Sedov-Taylor phase ever develops; consequently, phase-averaged models are not directly applicable to our case \citep[][see also \citealt{SerranoHernandezetal2025}. \citet{MartinezGonzalez2025b,MartinezGonzalez2025a,Kirchschlageretal2024b}]{MartinezGonzalezetal2018,MartinezGonzalezetal2019,MartinezGonzaletal2022}.

Currently, there are no available statistics on Blue Monsters or on the potential dependence of their properties on redshift. In our framework, the key parameters are the concentration and mass of the clusters. As more data become available, it will be possible to determine the conditions that lead to dust expulsion and, conversely, the conditions under which a substantial fraction of supernova-produced dust and gas are retained.

\section{Conclusions}
\label{sec:conclusions}

We combined net supernova dust yields from full 3-D hydrodynamical stellar cluster simulations with 3-D thin-shell blowout scalings to predict the dust-to-stellar mass ratio in young stellar clusters within super-early galaxies at $z\gtrsim10$. Mechanical blowout alone reduces $\xid$ by one to two orders of magnitude and reaches $\log\xid\approx-4$ over a wide range of radii, cloud-scale star formation efficiencies, and stellar cluster masses. The transition to very low $\xid$ follows gas concentration rather than metallicity, and remains effective at low $Z$. After weighting by the cluster mass function and a core-radius distribution, and without fine-tuning, the galaxy-integrated value $\log\langle\xid\rangle_{gal}$ lies in the same range as inferred values for spectroscopically-confirmed super-early galaxies commonly termed Blue Monsters. The most numerous clusters are also the most blowout-prone, hence their transparency is expected. A minority cluster population at the opposite extreme, consisting of very massive and extremely compact clusters with deep gravitational potentials, can suppress blowout and retain more dust and lead to a positive feedback scenario \citep{TenorioTagleetal2006,TenorioTagleetal2013,Silichetal2010,Amorinetal2009}. Since feedback in the form of radiation radiation pressure dominates only before the onset of supernovae \citep[][]{TenorioTagleetal2006,Grudicetal2018}, it cannot expel the bulk of the dust newly condensed in supernova ejecta. Supernova blowout therefore provides a natural channel for removing this dust into galactic halos.

\begin{acknowledgements}
The authors thank the anonymous referee for their helpful comments and suggestions that improved the quality of the paper. SJ acknowledges support by the Czech Ministry of Education, Youth and Sports, through the INTER-EXCELLENCE II program, project LUC24023, and by the institutional project RVO:67985815. This work is part of the collaboration ESTALLIDOS, supported by the Spanish research grants, PID2019-107408GB-C43 and PID2022-136598NB-C31 (ESTALLIDOS 8) from the Spanish Ministry of Science and Innovation.
The authors thankfully acknowledge the computer resources, technical expertise and support provided by the Laboratorio Nacional de Supercómputo del Sureste de México, SECIHTI member of the network of national laboratories.
\end{acknowledgements}

\bibliographystyle{aa}
\bibliography{blowout}

\end{document}